**World Scientific**
www.worldscientific.com

# A new approach to detecting gravitational waves via the coupling of gravity to the zero-point energy of the phonon modes of a superconductor*

Nader A. Inan

*School of Natural Sciences, University of California,
P. O. Box 2039, Merced, CA 95344, USA
ninan@ucmerced.edu*

The response of a superconductor to a gravitational wave is shown to obey a London-like constituent equation. The Cooper pairs are described by the Ginzburg–Landau free energy density embedded in curved spacetime. The lattice ions are modeled by quantum harmonic oscillators characterized by quasi-energy eigenvalues. This formulation is shown to predict a dynamical Casimir effect since the zero-point energy of the ionic lattice phonons is modulated by the gravitational wave. It is also shown that the response to a gravitational wave is far less for the Cooper pair density than for the ionic lattice. This predicts a "charge separation effect" which can be used to detect the passage of a gravitational wave.

## 1. Introduction

A "charge separation effect" has been predicted to occur in a superconductor in response to an incident gravitational wave.[1] This is fundamentally due to the Cooper pairs exhibiting a quantum rigidity (associated with the BCS energy gap) which makes them relatively nonresponsive to the gravitational wave. By contrast, the zero-point energy of the phonon modes of the ionic lattice dominates the response of the lattice to the gravitational wave. The difference in motion of the Cooper pairs (negatively charged) and the lattice ions (positively charged) results in the charge separation effect.[2] The resulting electric potential could be used as a means for detecting gravitational waves.

## 2. A Gravitational London-Like Constituent Equation

Recall that the London constituent equation for a superconductor is $\mathbf{J}_s = -\Lambda_L \mathbf{A}$, where the supercurrent $\mathbf{J}_s$ is induced by the vector potential $\mathbf{A}$. The London constant is $\Lambda_L = n_s e^2 / m_e$, where $n_s$ is the number density of Cooper pairs, $e$ the electron charge, and $m_e$ the electron mass. For gravitational waves, there is a similar relationship between $h_{ij}^{\tau\tau}$, the transverse-traceless strain field of a gravitational wave, and $T_{ij}^{\tau\tau}$, the transverse-traceless stress induced by the wave.[a]

The constituent equation is

$$T_{ij}^{\tau\tau} = -\mu_G h_{ij}^{\tau\tau}, \tag{1}$$

where $\mu_G$ is a positive, real constant with the dimensions of energy density. It can be described as a *gravitational shear modulus* which determines how much stress, $T_{ij}^{\tau\tau}$, is produced within a superconductor by an incident gravitational wave, $h_{ij}^{\tau\tau}$. Similar to the London equation in electromagnetism, the relationship in (1) follows as a direct result of the fact that particles within a superconductor (Cooper pairs and lattice ions) undergo *dissipationless* acceleration due to the gravitational wave field.

## 3. The Ginzburg–Landau (GL) Free Energy Density Embedded in Curved Spacetime

The response of the Cooper pair density to a gravitational wave can be formulated using the GL free energy density embedded in curved spacetime. First, recall the nonrelativistic GL free energy density in flat spacetime is

$$\mathscr{F} - \mathscr{F}_n = \frac{1}{2m} |(-i\hbar\nabla - q\mathbf{A})\psi|^2 + \alpha|\psi|^2 + \frac{\beta}{2}|\psi|^4, \tag{2}$$

where $m = 2m_e$, $q = 2e$, $\alpha$ and $\beta$ are phenomenological parameters, and $\psi$ is a complex order parameter.[4] To obtain a GL free energy coupled to gravity, we begin with the relativistic invariant, $g_{\mu\nu}\pi^\mu\pi^\nu = -m^2c^2$, where $\pi^\mu = (E/c, p^i)$ is the kinetic four-momentum. We expand the metric as $g_{\mu\nu} = \eta_{\mu\nu} + h_{\mu\nu}$, solve for the energy, and remain to first-order in the metric perturbation and to second-order in velocity. For gravitational waves in the far-field, the metric perturbation reduces to just the transverse-traceless part so that $h_{\mu 0} = 0$ and $h_{ij} = h_{ij}^{\tau\tau}$. In the nonrelativistic limit, it is also standard to drop the rest energy, $mc^2$, which gives

$$E = \frac{\pi^2}{2m} + \frac{h_{ij}^{\tau\tau}\pi^i\pi^j}{2m}, \tag{3}$$

where $\pi^i = -i\hbar\partial^i - qA^i$ for minimal coupling to electromagnetic fields. The Helmholtz free energy, $F = -k_B T \ln(Z)$, can be found using the partition function for a canonical ensemble, $Z = \sum_n \exp(-\beta E_n)$, with $E_n$ being the energy modes of the system, and $\beta = (k_B T)^{-1}$. For a superconductor, the Cooper pairs are essentially a Bose–Einstein condensate where the particles are in the zero-momentum ground state and therefore behave as effectively a single particle. In that case, the Helmholtz free energy is equivalent to the energy in (3).

---

[a]We use a metric constructed using the Helmholtz Decomposition theorem.[3] This formulation is gauge-invariant (to linear order in the metric) and isolates the radiative degrees of freedom as the transverse-traceless *part* of the metric perturbation, $h_{ij}^{\tau\tau}$.

Next we use a procedure similar to GL by introducing an order parameter, $\psi$, as well as a quadratic potential, $\alpha|\psi|^2$, and quartic self-coupling term, $\beta|\psi|^4$. To formulate the free energy density in *curved* spacetime, the coordinate volume can be expressed in terms of the proper volume as $dV = dV_{\text{proper}}/\sqrt{-g^{\tau\tau}}$, where $g^{\tau\tau}$ is the determinant of the metric. Then, the GL free energy density is

$$\mathscr{F}_{\text{GL}} = \sqrt{-g^{\tau\tau}} \left[ \frac{1}{2m}|\pi^i\psi|^2 + \frac{1}{2m}h_{ij}^{\tau\tau}(\pi^i\psi)^*(\pi^j\psi) + \alpha|\psi|^2 + \frac{\beta}{2}|\psi|^4 \right]. \tag{4}$$

The order parameter can be written as $\psi = \sqrt{n_s}\exp(\frac{i}{\hbar}\mathbf{p_0}\cdot\mathbf{r})$, where $\mathbf{p_0}$ is the momentum eigenvector. Since the Cooper pairs are in a zero-momentum eigenstate ($p_0 = 0$), then $\psi^*\psi = n_s$, which is a constant. Therefore, all derivatives vanish when using $\pi^i = -i\hbar\partial^i - qA^i$ in (4).

The Helmholtz free energy *density* is given by $d\mathscr{F} = -\mathbb{S}dT + T^{ij}dh_{ij}$, where $\mathbb{S}$ is the entropy density. It follows that the stress produced by a gravitational wave can be expressed in terms of the free energy density as

$$T^{ij} = \left(\frac{\partial\mathscr{F}}{\partial h_{ij}}\right)_T. \tag{5}$$

Using (4) leads to

$$\mu_{G(\text{CP})} = \frac{e^2 n_s}{m_e}(A^i)^2 + \alpha n_s + \frac{\beta}{2}n_s^2, \tag{6}$$

where "CP" represents Cooper pairs. For the zero-momentum eigenstate, the minimal coupling rule reduces to $mv^i = -qA^i$ which can be used in the first term of (6). We can also use the coherence length, $\xi = \hbar/\sqrt{2m_e|\alpha|}$, and the relation $n_s = -\alpha/\beta$, to write (6) as

$$\mu_{G(\text{CP})} = \frac{mn_s}{2}(v^i)^2 + \frac{\hbar^2 n_s}{4m_e\xi^2}. \tag{7}$$

The maximum kinetic energy which will preserve the superconducting state is determined by the BCS energy gap, $E_{\text{gap}} = \frac{7}{2}k_B T_c$, where $T_c$ is the critical temperature. Using values for niobium, we find that the first term in (7) dominates the second term and we have

$$\mu_{G(\text{CP})} \approx 4.9 \times 10^4 \, \text{J/m}^3. \tag{8}$$

To determine the *material* strain that would occur for the Cooper pair density, we use the *material* constituent equation, $T_{ij}^{\tau\tau} = -su_{ij}^{\tau\tau}$, where $s$ is the *material* shear modulus and $u_{ij}^{\tau\tau}$ is the strain tensor of the material. Equating this to (1) leads to

$$u_{ij}^{\tau\tau} = \frac{\mu_G}{s}h_{ij}^{\tau\tau}. \tag{9}$$

For niobium, the material shear modulus is $s \approx 3.8 \times 10^{10}$ J/m³. Therefore, (8) predicts that $u_{ij}^{\tau\tau}$ for the Cooper density is related to $h_{ij}^{\tau\tau}$ by a factor of $\frac{\mu_{G(\text{CP})}}{s} \approx 10^{-6}$. This means a gravitational wave on the order of $h_{ij}^{\tau\tau} \sim 10^{-17}$ will cause a material strain of only $u_{ij}^{\tau\tau} \sim 10^{-23}$ in the Cooper pair density, hence the Cooper pair is effectively nonresponsive.

## 4. The Debye Free Energy Embedded in Curved Spacetime

The lattice ions can be described by using the Debye model in curved spacetime. Using the approach of Ref. 5, we find that in the nonrelativistic limit, for gravitational waves in the far-field, the Hamiltonian is

$$\hat{H} = \frac{\hat{p}^2}{2m} + \frac{h_{ij}^{\tau\tau}\hat{p}^i\hat{p}^j}{2m} + \frac{K}{4}\hat{x}_i^2, \tag{10}$$

which includes a harmonic potential term to model the lattice ions as quantum harmonic oscillators.

For a gravitational wave with *periodic* time dependence, the Hamiltonian has a periodic behavior and thus applying Floquet's theorem leads to quasi-energy eigenvalues.[b] We use a plus-polarized gravitational wave propagating in the $z$-direction given by $h_{\oplus}(z,t) = A_{\oplus}\cos(kz)\sin(\omega t)$, where $A_{\oplus}$ is the amplitude of the gravitational wave. (Possible strain values are discussed in the conclusion section.) Also using a thin film approximation, $z \ll \lambda$, gives

$$E = \frac{1}{2m}(p_x^2 + p_y^2 + p_z^2) + \frac{K_i}{4}x_i^2 + \frac{1}{4\pi m}A_{\oplus}(p_x^2 - p_y^2). \tag{11}$$

We now use $n_x$, $n_y$ and $n_z$ to denote the occupation numbers of the modes with frequencies $\omega_x$, $\omega_y$, and $\omega_z$ in three directions, respectively. Applying separation of variables and summing over $N$ oscillators, with $n_\alpha$ being the number of phonons with frequency $\omega_\alpha$, we find that the quasi-energy eigenvalues in terms of phonon modes are

$$E_{(n_x,n_y,n_z)} = \hbar \sum_{\alpha}^{N} \omega_\alpha \left[ \sqrt{A_{\oplus}^+}\left(n_{x,\alpha} + \frac{1}{2}\right) + \sqrt{A_{\oplus}^-}\left(n_{y,\alpha} + \frac{1}{2}\right) + \left(n_{z,\alpha} + \frac{1}{2}\right)\right], \tag{12}$$

where $A_{\oplus}^{\pm} \equiv 1 \pm \frac{A_{\oplus}}{2\pi}$. Note that the presence of a gravitational wave breaks the spatial isotropy of the modes in each direction. In fact, because the gravitational wave couples to the zero-point energy of the oscillators ($\frac{1}{2}\hbar\omega$), this implies that the gravitational wave also breaks the isotropy of the *vacuum* energy of the ionic lattice as well. Since the gravitational wave is propagating in the $z$-direction, it modulates the frequencies in the $x$- and $y$-directions due to the squeezing/stretching of *space itself*. The factors, $\sqrt{A_{\oplus}^+}$ and $\sqrt{A_{\oplus}^-}$, are "gravitational modulation factors" which determine the modulation of the frequencies (and corresponding energies). Also, because the gravitational wave is dynamic, then it *dynamically* modulates the vacuum energy. (This is analogous to the mechanical oscillation of conducting plates which leads to the electromagnetic dynamical Casimir effect.)

---

[b]The method is described in Refs. 6 and 7.

The physical meaning of this dynamical Casimir effect is that an increase in the occupation number of the phonon modes of the lattice is predicted to occur in the presence of a gravitational wave. This effect is a quantum mechanical analog of a "Weber-bar effect" where the amplitude of the sound waves in the lattice grows in the classical limit due to the coupling of energy from the gravitational wave to the modes of the lattice.

Using the energy eigenvalues in (12), we obtain a partition function which leads to a Helmholtz free energy density in curved spacetime given by[c]

$$\mathscr{F}_D = \sqrt{-g_\oplus^{\tau\tau}} \left[ \frac{1}{2} n\hbar\omega \left( \sqrt{A_\oplus^+} + \sqrt{A_\oplus^-} + 1 \right) \right.$$

$$\left. - \frac{\pi}{6\hbar\beta^2 v} \left( \frac{1}{L_y L_z \sqrt{A_\oplus^+}} + \frac{1}{L_x L_z \sqrt{A_\oplus^-}} + \frac{1}{L_x L_y} \right) \right], \tag{13}$$

where $L_x$, $L_y$, $L_z$ are the dimensions of the superconductor.

Applying (5) and assuming $L_x \approx L_y \gg L_z$ for a thin, square superconducting film leads to

$$\mu_{G(\mathrm{LI})} \approx \frac{3\hbar\omega n}{2} - \frac{\pi}{3\hbar\beta^2 v L_x L_z}, \tag{14}$$

where "LI" represents the lattice ions. The first term can be referred to as the "zero-point energy density" since it originates from the zero-point energy of the system. Using the Debye frequency, $\omega_D$, which is the cut-off frequency in the Debye model, leads to a value of $2.8 \times 10^8 \, \mathrm{J/m^3}$ for niobium.

The second term in (14) can be referred to as the "sum of modes" term since it originates from summing over all the other modes of the system. Using a thickness for the superconducting film on the order of micrometers ($L_z \approx 10^{-6} \, \mathrm{m}$) and an edge on the order of centimeters ($L_x \approx 10^{-2} \, \mathrm{m}$), leads to a value of $9.0 \times 10^{-12} \, \mathrm{J/m^3}$. Therefore, we find that the "zero-point energy density" term dominates the "sum of modes" term and we simply have

$$\mu_{G(\mathrm{LI})} \approx 2.8 \times 10^8 \, \mathrm{J/m^3}. \tag{15}$$

Comparing this to (8), it is evident that the lattice ions are $\sim 10^4$ times more responsive to a gravitational wave than the Cooper pairs. This will lead to a relative strain between Cooper pairs and lattice ions in the presence of a gravitational wave.

---

[c]We are applying "periodic thermodynamics" as described by Refs. 8 and 9.

## 5. Conclusion

The relative strain between lattice ions and Cooper pairs (induced by the gravitational wave) is $U_{ij}^{\tau\tau} \equiv u_{ij(\mathrm{LI})}^{\tau\tau} - u_{ij(\mathrm{CP})}^{\tau\tau}$. Using (9) gives

$$U_{ij}^{\tau\tau} = \frac{\mu_{G(\mathrm{LI})} - \mu_{G(\mathrm{CP})}}{s} h_{ij}^{\tau\tau}. \tag{16}$$

Inserting the results from (8) and (15), and making use of $s \approx 3.8 \times 10^{10} \, \mathrm{J/m}^3$ for niobium, gives

$$U_{ij}^{\tau\tau} \approx (10^{-2}) h_{ij}^{\tau\tau}. \tag{17}$$

Hence, there is a relative strain between the Cooper pairs and lattice ions that is predicted to occur on the order of 1% of the gravitational wave strain field. This is a result of the difference between $\mu_{G(\mathrm{LI})}$ and $\mu_{G(\mathrm{CP})}$ in (16). Note that this difference has a *quantum mechanical* origin, namely, the difference between the BCS energy gap (preserving the Cooper pairs) and the zero-point energy of the phonon modes of the ionic lattice. Therefore, this is a macroscopic *quantum* effect which has no classical analog.

Since Cooper pairs and lattice ions are oppositely charged, then the relative strain predicted in (17) is essentially a "charge separation effect" caused by the gravitational wave. This implies that the gravitational wave will induce an electric potential in the superconductor which could be experimentally detected. Experiments with a high-$Q$ niobium resonant mass gravitational radiation antenna have already achieved a strain sensitivity of $10^{-19}$.[10] Therefore, using (17) implies that a gravitational wave strain of the order of $\sim 10^{-17}$ could be detected using such an apparatus.

Finally, we give some bounds on the frequencies for which a superconductor will respond to a gravitational wave with the charge separation effect described above. The energy deposited in the superconductor by the gravitational wave is given by $E = \hbar\omega$. To preserve the superconducting state of the system, this energy must not exceed the BCS energy gap, $E_{\mathrm{gap}} = \frac{7}{2} k_B T_C$, where $k_B$ is the Boltzmann constant and $T_C$ is the critical temperature. For niobium, $T_C = 9.3 \, \mathrm{K}$. This leads to an upper bound of $\omega \sim 10^{12} \, \mathrm{Hz}$ for the allowed frequencies of the incident gravitational wave. For a lower bound, it is predicted that the charge separation effect is valid down to the DC limit.

Note that these bounds for the frequencies demonstrate that the charge separation effect is *not* an antenna-like effect which would ordinarily require the dimensions of the antenna to be well above the wavelength of the detected wave. Rather, the effect is a *Meissner-like effect* where the external gravitational wave is *expelled* from a superconductor much like the magnetic field is expelled from a superconductor in the standard Meissner effect, which also occurs in the DC limit. Other alternative methods for detecting gravitational waves can also be found in Refs. 11–14.

**Note Added in Proof**

A possible objection concerning the charge separation effect is that it violates the equivalence principle which predicts that all particles will freely fall exactly the same way in a gravitational field, independent of their composition and their electrical charge (including the sign of their charge). Therefore, charges will not separate from one another in free fall, and bulk matter should stay electrically neutral in response to gravitational fields.

However, the equivalence principle does not apply to a system where forces other than gravity are acting on the masses. Specifically, the Cooper pair density in a superconductor experiences a quantum rigidity characterized by the complex order parameter. This introduces a type of restoring force on the scale of the coherence length of the superconductor as shown in (7). Likewise, the lattice ions experience inter-atomic Coulomb forces which are characterized by the harmonic potential term found in the Hamiltonian for an ensemble of quantum harmonic oscillators (10). Therefore, the charge separation effect does not violate the equivalence principle since the masses (Cooper pairs and lattice ions) are not freely falling in the presence of the gravitational field. In essence, the gravitational field acts as any other external force on the system, but the Cooper pair density and the ionic lattice each exhibit a different restoring force as demonstrated by the difference in values between (8) and (15). Inserting these values into (1) implies a different response by the Cooper pairs and lattice ions, hence leading to the charge separation effect.

**References**

1. S. Minter, K. Wegter-McNelly and R. Chiao, *Physica E* **42** (2010) 234, arXiv: 0903.0661.
2. N. A. Inan, J. J. Thompson and R. Y. Chiao, *Fortschr. Phys.* **65** (2017) 1600066, doi: 10.1002/prop.201600066.
3. É. Flanagan and S. Hughes, *New J. Phys.* **7** (2005) 204.
4. M. Tinkham, *Introduction to Superconductivity*, 2nd edn. (McGraw Hill, New York, 1976).
5. B. DeWitt, *Phys. Rev. Lett.* **16** (1966) 1092.
6. Y. Zel'dovich, *Sov. Phys.-JETP* **24** (1967) 1492.
7. H. Sambe, *Phys. Rev. A* **7** (1973) 2203.
8. W. Kohn, *J. Stat. Phys.* **103** (2001) 417.
9. M. Langemeyer and M. Holthaus, *Phys. Rev. E* **89** (2014) 01201.
10. D. G. Blair *et al.*, *Phys. Rev. Lett.* **74** (1995) 1908.
11. C. Sabin *et al.*, *New J. Phys.* **16** (2014) 085003, arXiv:1402.7009 [quant-ph].
12. A. Arvanitaki and A. Geraci, *Phys. Rev. Lett.* **110** (2013) 071105, arXiv:1207.5320 [gr-qc].
13. P. Jones *et al.*, *Phys. Rev. D* **95** (2017) 065010, arXiv:1610.02973 [gr-qc].
14. P. Jones and D. Singleton, *Int. J. Mod. Phys. D* **24** (2015) 1544017, arXiv:1505.04843 [gr-qc].